\documentclass[prd,twocolumn,aps]{revtex4}

\newcommand{\beq}{\begin{equation}}  
\newcommand{\eeq}{\end{equation}}
\newcommand{\bea}{\begin{eqnarray}}
\newcommand{\eea}{\end{eqnarray}}

\usepackage{graphicx}

\begin{document}

\title{Yukawa sector in non-supersymmetric renormalizable SO(10)}

\author{
Borut Bajc$^{(1)}$,  
Alejandra Melfo$^{(2,3)}$, 
Goran Senjanovi\'c$^{(3)}$ 
and Francesco Vissani$^{(4)}$}
\affiliation{$^{(1)}$ {\it J.\ Stefan Institute, 1001 Ljubljana, Slovenia}}
\affiliation{$^{(2)}$ {\it Centro de F\'{\i}sica Fundamental, 
Universidad de Los Andes, M\'erida, Venezuela}}
\affiliation{$^{(3)}${\it International Center for Theoretical Physics, 
Trieste, Italy}}
\affiliation{$^{(4)}${\it INFN, Laboratori Nazionali 
del Gran Sasso, Theory Group, Italy}}

\begin{abstract}
We discuss the ordinary, non-supersymmetric  SO(10) as a theory of 
fermion masses and mixings. We construct two minimal versions of the 
Yukawa sector based on
${\bf\overline{126}_H}$ and either $\bf{10_H}$ or $\bf{120_H}$.  
The latter case is of particular interest since it connects
the absolute neutrino mass scale with the size of the  
atmospheric mixing angle $\theta_A$. It also relates the smallness of 
$V_{cb}$ with the largeness of $\theta_A$. These results are based on 
the analytic study of the second and third generations. Furthermore, 
we discuss the structure of the light Higgs and the role 
of the Peccei-Quinn symmetry for dark matter and the predictivity 
of the theory.
\end{abstract}


\maketitle

\section{Introduction}

SO(10) grand unified theory \cite{Fritzsch:nn} is probably the best
motivated candidate for the unification of the strong and electro-weak
interactions. It unifies the family of fermions; it includes the 
SU(4)$_C$ quark-lepton symmetry \cite{Pati:1974yy} and the left-right (LR) 
symmetry \cite{leftright} in the form of charge conjugation as a finite
gauge symmetry; it predicts the existence of right-handed neutrinos and
through the see-saw mechanism \cite{Minkowski:1977sc} offers an appealing 
explanation for the smallness of neutrino masses.  Due to the success of
supersymmetric unification, and the use of supersymmetry in controlling
the gauge hierarchy, most of the attention in recent years has focused on
the supersymmetric version of SO(10). However, supersymmetry may not be
there. After all, it controls the Higgs gauge hierarchy, but not the
cosmological constant. The long standing failure of understanding the
smallness of the cosmological constant suggests that the unwelcome
fine-tuning may be necessary. Our fine-tuned world can be viewed, in the 
landscape picture simply as a selection criterion among the 
large number of degenerate string vacua. Or it could be that 
the cosmological evolution of the universe selects a
light Higgs doublet \cite{Dvali:2003br}. 
If so, the main motivation behind the low-energy supersymmetry would 
be gone. However one could still hope that the issue might be settled by
the internal consistency and the predictions of a well defined grand 
unified theory. In other words, it is possible that simplicity and 
minimality at the GUT scale requires a specific low-energy theory. 

What about grand unification without supersymmetry?  At first glance, 
one may worry about the unification of gauge couplings in this case.
Certainly, in the minimal SU(5) theory, the gauge couplings do not 
unify without low-energy supersymmetry. What happens is the following: 
the colour and weak gauge couplings meet at around $10^{16}$ GeV, an 
ideal scale from the point of view of the proton stability and 
perturbativity (i.e., sufficiently below $M_{Planck}$). The problem 
is the U(1) coupling. Without supersymmetry it meets the SU(2)$_L$ 
coupling at around $10^{13}$ GeV \cite{Langacker:1991an}; with low-energy 
supersymmetry the one-step unification works as is well known 
\cite{susyunif}.  

On the other hand, the fact that neutrinos are massive indicates 
strongly that SU(5) is not the right grand unified theory: it simply 
requires too many disconnected parameters in the Yukawa sector 
\cite{Aulakh:2003kg,Senjanovic:2005sf}. The SO(10) theory is favored 
by the neutrino oscillation data. Most interestingly, SO(10) 
needs no supersymmetry for a successful unification of gauge couplings. 
On the contrary, the failure of ordinary SU(5) tells us that in the 
absence of supersymmetry there is necessarily an intermediate scale 
such as the left-right symmetry breaking scale $M_R$. Namely, in this 
case the SU(2)$_L$ and SU(3)$_C$ couplings run as in the standard model 
or with a tiny change depending whether or not there are additional 
Higgs multiplets at $M_R$ (recall that the Higgs contribution to the 
running is small). However, the U(1)  coupling is strongly affected by 
the embedding in SU(2)$_R$ above $M_R$. The large contributions of the 
right-handed gauge bosons makes the U(1) coupling increase much slower 
and helps it meet the other two couplings at the same point. The scale 
$M_R$ typically lies between $10^{10}$ GeV and $10^{14}$ GeV (see for 
example \cite{Deshpande:1992em,Acampora:1994rh} and references therein), 
which fits very nicely with the see-saw mechanism. Now, having no 
supersymmetry implies the loss of a dark matter candidate. One may be 
even willing to introduce an additional symmetry in order to achieve this. 
In this case it should be stressed that SO(10) provides a framework for 
the axionic dark matter:  all one needs is a Peccei-Quinn 
\cite{Peccei:1977hh} symmetry U(1)$_{PQ}$ which simultaneously 
solves the strong CP problem. 

This seems to us more than sufficient motivation to carefully study 
ordinary non-supersymmetric SO(10). What is missing in this program is 
the construction of a well defined predictive theory with the realistic 
fermionic spectrum. This paper is devoted precisely to this task. 
 
In particular, the search of the minimal realistic Yukawa sector is 
a burning question. In the absence of higher dimensional operators at 
least two Higgs multiplets with the corresponding Yukawa matrices are 
needed, otherwise there would be no mixings. The Yukawa Higgs sector 
can contain $\bf{10_H}$, $\bf{120_H}$ and ${\bf\overline{126}_H}$ 
representations, since 

\beq 
{\bf 16\times 16=10+120+126}\;.
\eeq

One version of the theory with only $\bf{10_H}$ and 
${\bf \overline{126}_H}$ was studied in great detail in the case of 
low-energy supersymmetry  \cite{Bajc:2004xe,Aulakh:2002zr,Fukuyama:2004ps}. In spite of having 
a small number of parameters it seems to be consistent with all the data 
\cite{Matsuda:2000zp,Matsuda:2001bg,Goh:2003sy,Babu:2005ia,Bertolini:2005qb}. 
For the type II seesaw it predicts furthermore the $1-3$ leptonic mixing 
angle not far from its experimental limit: $|U_{e3}| > 0.1$ 
\cite{Goh:2003sy,Bertolini:2005qb} and it offers an interesting connection 
between $b-\tau$ unification and the large atmospheric mixing angle 
\cite{Bajc:2002iw,Bajc:2004fj}. Last but not least, it also predicts exact 
R-parity at low energies \cite{rparity,Aulakh:1997ba} leading to the LSP 
as a candidate for the dark matter. 
 
Thus, a first obvious possibility in ordinary SO(10) is to address the 
model with ${\bf 10_H} + {\bf \overline{126}_H}$, and to see whether 
or not it can continue to be realistic. 

There is another interesting alternative: $\bf{120_H}$ and 
${\bf\overline{126}_H}$. We limit ourselves to the analytic study, 
which requires ignoring the  effects of the first generation. 
The main result of this model is the correlation of neutrino 
masses with the value of the atmospheric mixing angle, true 
both in Type I and Type II seesaw mechanisms, or in a general case 
when both contribute to the neutrino masses. Furthermore the large 
atmospheric mixing angle fits naturally with the small $V_{cb}$ mixing. 

Both cases require complexifying the Higgs fields $10_H$ or $120_H$. 
This in turn calls for a PQ symmetry, in any case useful for a 
dark matter. 

At first glance it seems that the Yukawa sector in a non-supersymmetric 
theory does not differ from the supersymmetric version and thus there 
is nothing new to say. There are however several subtle differences: 
1) the running of the couplings (gauge and Yukawa) is changed, so are 
the inputs for a numerical evaluation at $M_{GUT}$; 2) there are 
necessarily intermediate scales; 3) if no new symmetries are invoked, 
all the SO(10) representations that are not complex like ${\bf 16}$ or 
${\bf 126}$ are real; 4) there are radiative corrections to the Yukawa 
sector, that should in principle be taken into account. All these 
points will be discussed below. 

\section{The minimal Yukawa sector}

In this work, we stick to the renormalizable version of the see-saw 
mechanism (for alternatives using a radiatively-induced see-saw, 
see\cite{Bajc:2005aq}), which makes the representation 
${\bf \overline{126}_H}$ indispensable, since it breaks the SU(2)$_R$ 
group and gives a see-saw neutrino mass. By itself it gives no fermionic 
mixing, so it does not suffice. The realistic fermionic spectrum 
requires adding either $\bf{10_H}$ or $\bf{120_H}$. As promised in 
the Introduction, we will go carefully through both possibilities. 

Before starting out, it is convenient to decompose the Higgs fields 
under the SU(2)$_L\times$ SU(2)$_R\times$ SU(4)$_C$ Pati-Salam (PS) group:

\begin{eqnarray}
{\bf 10} &=& (2,2,1) + (1,1,6)  \nonumber \\
{\bf\overline{126}} &=& (1,3,10) + (3,1,\overline{10}) + (2,2,15) + (1,1,6) \\
{\bf 120} &=& (1,3,6) + (3,1,6) + (2,2,15) + (2,2,1) + (1,1,20)\nonumber
\end{eqnarray}

\noindent
As is well known, the ${\bf 126_H}$ provides mass terms for 
right-handed and left-handed neutrinos:

\begin{equation}
M_{\nu_R}=  \langle 1,3,10 \rangle  \,Y_{126},\ \ 
M_{\nu_L}=  \langle 3,1,\overline{10}\rangle  \, Y_{126} 
\label{mlr}
\end{equation}

\noindent
which means that one has both type I and type II seesaw:

\beq 
M_N= -M_{\nu_D}M_{\nu_R}^{-1}M_{\nu_D}+M_{\nu_L}\;
\label{mnu}
\eeq

\noindent
In the type I case it is the large vev of $(1,3,10)$ that provides the 
masses of right-handed neutrino whereas in the type II case, the 
left-handed triplet provides directly light neutrino masses through a 
small vev \cite{Lazarides:1980nt,Mohapatra:1980yp}. 
The disentangling of the two contributions is in general hard. 
 
\subsection{ Model I: ${\bf \overline{126}_H}$  + $\bf{10_H}$  }

In this case the most general Yukawa interaction is (schematically)

\begin{equation}
\label{yukawa}
{\cal L}_Y={\bf 16_F}\left({\bf 10_H} Y_{10}+
{\bf \overline{126}_H} Y_{126}\right){\bf
16_F}+h.c.\;.
\end{equation}

\noindent
where $Y_{10}$ and $Y_{126}$ are symmetric matrices in the generation space. 
With this one obtains relations for the Dirac fermion masses 

\begin{eqnarray}
M_D=M_1+M_0&,\;&M_U=c_1M_1+c_0 M_0\;\;,\nonumber\\
M_E=-3M_1+  M_0& , \;&M_{\nu_D}=-3c_1M_1+ c_0 M_0,
\label{mod1}
\end{eqnarray}

\noindent
where we have defined

\begin{eqnarray}
M_1&=& \langle 2,2,15 \rangle_{126}^d  \, Y_{126}\;, \nonumber \\
M_0&=& \langle 2,2,1 \rangle_{10}^d  \, Y_{10} \; ,
\label{m1m0}
\end{eqnarray}

\noindent
and

\begin{equation}
c_0=\frac{\langle 2,2,1\rangle_{10}^u }{ 
\langle 2,2,1\rangle_{10}^d } \, , \;
 c_1=\frac{\langle 2,2,15\rangle_{126}^u }{
\langle 2,2,15\rangle_{126}^d }\;.
\label{c1c0}
\end{equation}

\noindent
These equations, together with ~(\ref{mlr}) and  (\ref{mnu}), 
are the starting point for the analysis 
\cite{Babu:1992ia} of the fermion spectrum. 

\subsubsection{Who is the light Higgs?\label{whois}}

With the minimal fine-tuning the light Higgs is in general a 
mixture of, among others,  (2,2,1) of ${\bf 10_H}$ and (2,2,15) of 
${\bf\overline{126}_H}$. This happens at least due to the large (1,3,10) 
vev in the term $({\bf \overline{126}_H})^2 \, 
{\bf \overline{126}_H}^\dagger \,{\bf 10_H}$.

In any case, their mixings require the breaking of SU(4)$_C$ symmetry at 
a scale $M_{PS}$, and it is thus controlled by the ratio $ M_{PS}/M$, 
where $M$ corresponds to the mass of the heavy doublets. Thus, if 
$M\simeq M_{GUT}$, and $M_{PS} \ll M_{GUT}$, this would not work; 
we come to the conclusion that one needs 
to tune-down somewhat $M$. Tuning down of the $(2,2,1)$ mass 
cannot have much impact on the unification constraints, but $(2,2,15)$ 
is a large field and could in principle cause trouble. However, its 
contribution is actually tiny, since the differences in the 
corresponding $\beta$-function coefficients $(b_2 - b_3)_{(2,2,15)}$ and 
$(b_1 - b_2)_{(2,2,15)}$ (in the usual notation) are very small. It should 
nevertheless be taken into account when studying unification constraints.
 
\subsubsection{The simplest version: real ${\bf 10_H}$ }

If ${\bf 10_H}$ is real, then there is just one SU(2)$_L$ doublet 
in $(2,2,1)$ and thus $|\langle 2,2,1\rangle_{10}^u|= 
|\langle2,2,1\rangle_{10}^d|$, namely $|c_0|=1$. 
The parameter space is thus smaller. 
Here we show that, in the two generation (second and third) case 
with real parameters, such a constraint leads to a contradiction 
with the data. In the physically sensible approximation 
$\theta_q= V_{cb}=0$ we find 

\begin{equation}
c_0=\frac{m_c(m_\tau-m_b)-m_t(m_\mu-m_s)}{m_sm_\tau-m_\mu m_b}
\approx \frac{m_t}{m_b}\;,
\end{equation}

\noindent
clearly much bigger than $1$.

This conclusion is subject to the uncertainties of the full 
three-generation case. Although strictly speaking this simple model
cannot be ruled out yet, there is an indication that a more complicated 
scenario should be considered.

\subsubsection{The next step: complex ${\bf 10_H}$}

If the model with real $\bf{10_H}$ does fail eventually, one could 
simply complexify it. This of course introduces new Yukawa couplings 
which makes the theory less predictive. Certain predictions may remain, 
though, such as the automatic connection between $b-\tau$ unification 
and large atmospheric mixing angle in the type II see-saw. This is true
independently of the number of 10 dimensional Higgs representations, 
since $\bf{10_H}$ cannot distinguish down quarks from charged leptons. 
{}From $M_{\nu_L}\propto Y_{126}\;,$ one has $M_{\nu_L}\propto M_D-M_E$.

It is a simple exercise to establish the above mentioned connection
between $|m_b|\approx |m_\tau|$ and large $\theta_{atm}$; for details 
see \cite{Bajc:2002iw,Bajc:2004fj}. This fact has inspired the careful 
study of the analogous supersymmetric version where $m_\tau \simeq m_b$ 
at the GUT scale works rather well. In the non-supersymmetric theory, 
$b-\tau$ unification fails badly, $m_\tau \sim  2 m_b$ \cite{Arason:1992eb}. 
The realistic theory will require a Type I seesaw, or an admixture of 
both possibilities.

\subsubsection{Axions and the dark matter of the universe}
\label{axion}

A complex $\bf{10_H}$ means, as we said, an extra set of Yukawa 
couplings. At the same time this non-supersymmetric theory cannot 
account for the dark matter of the universe, since there are no 
cosmologically stable neutral particles and, as is well known, light 
neutrinos cannot too. It is then rather suggestive to profit from the 
complex $\bf{10_H}$ and impose the U(1)$_{PQ}$ Peccei-Quinn symmetry:

\beq
{\bf 16}\to e^{i\alpha} {\bf 16}\;,\;
{\bf 10}\to e^{-2 i \alpha}{\bf 10}\;,\;
{\bf\overline{126}}\to e^{-2 i \alpha}{\bf\overline{126}}\;,
\eeq

\noindent
with all other fields neutral. The Yukawa structure has the form 
(\ref{yukawa}) with  ${\bf 10_H}$ now complex. This resolves the 
inconsistency in fermion masses and mixings discussed above, and gives 
the axion as a dark matter candidate as a bonus \cite{axions}.

The neutrality of the other Higgs fields under U(1)$_{PQ}$ emerges from 
the requirement of minimality of the Higgs sector that we wish to stick 
to. Namely, ${\bf \overline{126}_H}$ is a complex representation and 
${\bf 10_H}$ had to be complexified in order to achieve realistic 
fermion mass matrices and to have U(1)$_{PQ}$. It is desirable that the 
U(1)$_{PQ}$ be broken by a nonzero $\langle{\bf \overline{126}_H}\rangle$, 
i.e. the scale of SU(2)$_R$ breaking and right-handed neutrino masses 
\cite{Mohapatra:1982tc}, otherwise $\bf{10_H}$ would do it an 
give $M_{PQ} \approx M_W$, which is ruled out by experiment. Actually, 
astrophysical and cosmological limits prefer $M_{PQ}$ in the window 
$10^{10} - 10^{13}$ GeV   \cite{Kim:1986ax}.

Now, a single ${\bf\overline{126}_H}$ just trades the original 
Peccei-Quinn charge for a linear combination of U(1)$_{PQ}$, 
$T_{3R}$ and $B-L$ \cite{Mohapatra:1982tc,Holman:1982tb}. 
Thus in order to break this combination and provide the Goldstone 
boson an additional Higgs multiplet is needed. One choice is to add 
another ${\bf \overline{126}_H}$ and decouple it from fermions, 
since it must necessarily have a different PQ charge 
\cite{Mohapatra:1982tc}. An alternative is to use a (complex) GUT 
scale Higgs as considered for SU(5) by \cite{Wise:1981ry}, with 
$M_{PQ} \simeq M_{GUT}$, which however implies too much dark matter 
or some amount of fine-tuning. 

Of course, the Peccei-Quinn symmetry does more than just providing the 
dark matter candidate: it solves the strong CP problem and predicts the 
vanishing $\bar\theta$. The reader may object to worrying about the strong 
CP and not the Higgs mass hierarchy problem; after all, they are both  
problems of fine-tuning. Actually, the strong CP problem is not even a 
problem in the standard model, at least not in the technical sense 
\cite{Ellis:1978hq}. Namely, although divergent, in the standard model 
$\bar\theta$ is much smaller than the experimental limit: $\bar\theta\ll 
10^{-10}$ for any reasonable value of the cutoff $\Lambda$, e.g. 
$\bar\theta\approx 10^{-19}$ for $\Lambda=M_{Planck}$. The physical 
question is really the value of $\bar\theta$. Peccei-Quinn symmetry 
fixes this arbitrary parameter of the SM. The situation 
with supersymmetry and the Higgs mass is opposite. Low energy SUSY helps 
keep Higgs mass small in perturbation theory, but fails completely in 
predicting it. If we do not worry about the naturalness we can do without 
supersymmetry. On the other hand, if we wish to predict the electron 
dipole moment of the neutron, U(1)$_{PQ}$ is a must, unless we employ 
the spontaneous breaking of $P$ or $CP$ in order to control $\bar \theta$
\cite{Beg:1978mt,Mohapatra:1978fy,Georgi:1978xz}.

\subsection{Model II: ${\bf \overline{126}_H}$  + $\bf{120_H}$ }
 
Instead of $\bf{10_H}$ one could use $\bf{120_H}$. Since $Y_{120}$ 
is antisymmetric, this means only 3 new complex couplings on top of 
$Y_{126}$. 

$\bf{120_H}$ as an addition to $\bf{10_H}$ was partially studied some 
time ago, for charged fermions only \cite{Matsuda:2000zp}. Recently it 
was readdressed \cite{Bajc:2005aq} in the context of the radiative seesaw 
mechanism \cite{Witten:1979nr}. The analytic study for the second and 
third generation gives $b-\tau$ unification, and small quark and large 
leptonic mixing angle.  We recall that this  model 
is tailor-fit for the strongly split 
supersymmetry \cite{Arkani-Hamed:2004fb} with light gauginos 
and higgsinos and very heavy scalars \cite{Bajc:2004hr}.

The analysis here is quite similar to the case of ${\bf 10_H}$ and 
${\bf 120_H}$ due to the fact that $\bf{10_H}$ and 
${\bf\overline{126}_H}$ have symmetric Yukawas. There is an important 
difference though, since $(2,2,1)_{10}$ is traded for 
$(2,2,15)_{126}$.

The Dirac mass matrices at the grand unification scale take the form 
 
The Dirac mass matrices at the grand unification scale take the form 
\begin{eqnarray}
M_D=M_1+M_2&\;\;,\;\;&M_U=c_1M_1+c_2M_2\;\;, \\
M_E=-3M_1+c_3M_2&\;\;,\;\;&M_{\nu_D}=-3c_1M_1+c_4M_2\; \nonumber
\end{eqnarray}
\noindent
where $M_1$ and $c_1$ are defined in (\ref{m1m0}),(\ref{c1c0}), and:
\begin{eqnarray}
M_2&=&Y_{120}\left(\langle 2,2,1 \rangle_{120}^d\ +
\langle 2,2,15 \rangle_{120}^d \right)\;, \nonumber \\
c_2 &=& \frac{\langle 2,2,1 \rangle_{120}^u +
\langle 2,2,15 \rangle_{120}^u }{\langle 2,2,1 \rangle_{120}^d +
\langle 2,2,15 \rangle_{120}^d }\;, \nonumber  \\
c_3&=&{\langle 2,2,1\rangle_{120}^d-3
\langle 2,2,15\rangle_{120}^d\over 
\langle 2,2,1\rangle_{120}^d+
\langle 2,2,15\rangle_{120}^d}\;, \nonumber \\
c_4&=&{\langle 2,2,1\rangle_{120}^u-3
\langle 2,2,15\rangle_{120}^u\over 
\langle 2,2,1\rangle_{120}^d+
\langle 2,2,15\rangle_{120}^d}\;.
\label{lec}
\end{eqnarray}
The case of real $120_H$ reduces to similar constraints already 
encountered in the real case of model I (see also \cite{Nath:2001uw} for an analysis using SU(5)
decomposition).
For real bidoublets 
the definitions (\ref{lec}) constrain all three $c_i$ to 
a same order of magnitude. But this contradicts the requirements 
for small second generation masses of charged leptons ($c_3\approx 3$) 
and of up quarks ($c_2\approx m_t/m_b$). In other words, similarly 
as in model I, there is a need to complexify the Higgs fields. 
This is again best achieved by introducing a U(1)$_{PQ}$ global 
symmetry, which provides as a byproduct a dark matter candidate.

The type I seesaw contribution due to right-handed neutrinos 
gives the light neutrino mass matrix

\begin{equation}
M_N^I=-M_{\nu_D}^TM_{\nu_R}^{-1}M_{\nu_D}\propto 
9c_1^2M_1-c_4^2M_2M_1^{-1}M_2
\label{se1}
\end{equation}

\noindent
whereas the type II contribution reads

\begin{equation}
M_N^{II}\propto M_1.
\label{se2}
\end{equation}

\subsubsection{Two generations case: analysis and predictions}

In spite of only two Yukawa matrices, the above system of equations
is rather complicated and requires painstaking numerical studies. 
Before plunging in this computational tedium, 
it is certainly useful if not indispensable to get a physical 
insight through analytical arguments. 
Following the successful approach that was adopted by us before
(for the case of a supersymmetric SO(10) theory with ${\bf 10_H}$ and 
${\bf \overline{126}_H}$ \cite{Bajc:2002iw,Bajc:2004fj}) 
we focus here on the study of the second and third 
generations, with the natural expectation
that the effects of the first generation can be treated 
as a perturbation. 

In the basis where $M_1$ is diagonal, real and non-negative:

\begin{equation}
M_1\propto 
\left(
\begin{array}{cc}
\sin^2\theta & 0 \\
0 & \cos^2\theta
\label{m1}
\end{array}
\right)
\label{deag}
\end{equation}

\noindent
the most general charged fermion matrix 
can be written as:

\begin{equation}
M_f=\mu_f 
\left(
\begin{array}{cc}
\sin^2\theta & i (\sin\theta \cos\theta+\epsilon_f) \\
-i (\sin\theta \cos\theta+\epsilon_f) & \cos^2\theta
\end{array}
\right)\;,
\label{eqq2}
\end{equation}

\noindent
where $f=D,U,E$ stands for charged fermions and $\epsilon_f$ vanishes 
for negligible second generation masses. In other words 
$|\epsilon_f|\propto m_2^f/m_3^f$ (see below). Furthermore 
the real parameter $\mu_f$ sets the third generation mass scale, 
made explicit below. 

We determine next the unitary matrices $L_f$ and $R_f$, which diagonalize 
$M_f$ in the physically relevant approximation of small $|\epsilon_f|$. 
We obtain

\begin{equation}
M=R_f^t\cdot {\rm diag}\{-\mu_f \epsilon_f \sin 2\theta,\mu_f(1+
\epsilon_f \sin 2\theta)\}\cdot L_f 
+ {\cal O}(|\epsilon^2_f|)
\label{decom}
\end{equation}

\noindent
where

\begin{equation}
L_f=
\left(
\begin{array}{cc}
1 & -i \cos 2\theta\ \overline{\epsilon}_f \\
-i \cos 2\theta\ \epsilon_f & 1
\end{array}
\right)
\left(
\begin{array}{cc}
\cos\theta & -i \sin\theta \\
-i \sin\theta & \cos\theta
\end{array}
\right)
\label{quaglia}
\end{equation}

\noindent
and $R_f$ is given by the same expression with $i\to -i$. From 
eq.~(\ref{decom}), at the leading order in  $|\epsilon|$ we get:

\begin{eqnarray}
|\mu_f| & = & m^f_3\;,\\
\sin{2\theta}|\epsilon_f|&= &m^f_2/m^f_3\;.
\end{eqnarray}

\noindent
The phases of the three $\epsilon_f$ parameters are not determined, 
while the meaning of the angle $\theta$ will 
be clear in a moment. Now we are in the position to 
state the three predictions of this theory regarding 1) neutrino masses,  
2) the relation between bottom and tau masses, and 3) the quark 
mixing angle $V_{cb}$.

1) We begin with the predictions regarding neutrino masses.
By using eqs.~(\ref{se1}) and (\ref{se2}) and an explicit 
form of the $2\times 2$ matrices one concludes that type I 
and type II seesaw lead to the same structure  

\begin{equation}
M_N^I\propto M_N^{II}\propto M_1 
\end{equation}

\noindent
In the selected basis the neutrino mass matrix is diagonal.
We see that the angle $\theta$ has to be identified with the leptonic 
(atmospheric) mixing angle $\theta_A$ up to terms of the order 
of $|\epsilon_E|\approx m_\mu/m_\tau$. 
For the neutrino masses we obtain from (\ref{m1})

\begin{equation}
\label{dmsm}
\frac{m_3^2-m_2^2}{m_3^2+m_2^2}=
\frac{\cos2\theta_A}{1-\sin^22\theta_A/2} + {\cal O}(|\epsilon|)
\end{equation}  

\noindent
This equation points to an intriguing correlation: the degeneracy 
of neutrino masses is measured by the maximality 
of the atmospheric mixing angle.
What about numerical predictions? Clearly, without including the
effects of the first generation and the impact of the running from the
GUT to the weak scale, no precise determination can be made.
It may be illustrative though to give an estimate in case this formula
were to remain approximately valid.    
The value of $m_2$ could not then be too small: e.g., with the 
value $\Delta m^2_A=|m_3^2-m_2^2|\approx 2.5\cdot 10^{-3}$ eV$^2$
and the 99 \% CL limit $\theta_A=45^\circ\pm 9^\circ$  
from \cite{Strumia:2005tc} one would get $m_2>30$~meV.
On the other hand, there is an upper limit 
from cosmology and neutrinoless double beta decay,
which (depending on the selected data and analysis) 
varies from $0.14$ eV to $0.5$~eV, see again \cite{Strumia:2005tc}. 
Clearly, the larger the limit, the 
closer one can be to $\theta_A=45^\circ$. The analysis in 3) below 
suggests though that $\theta_A$ should be as far as possible from the 
maximal value, i.e.\ that neutrinos should be as 
hierarchical as allowed by (\ref{dmsm}).

2) The second prediction regards the ratio of tau and bottom mass at 
the GUT scale: 

\begin{equation}
\frac{m_\tau}{m_b}=3 + 3 \sin 2\theta_A\ {\rm Re}[\epsilon_E-\epsilon_D]
+{\cal O}(|\epsilon^2|)
\end{equation}

\noindent
At first glance, this appears to kill the model; after all, 
the extrapolation in the standard model leads to expect 
$m_\tau\approx 2 m_b$. However, it is not
possible to exclude that several effects modify this 
conclusion and avoid a flat contradiction with data (although 
we would in any case expect that $m_b$ comes out 
as small as possible).
In particular, we note that with a suitable choice of phases the  
corrections order $\epsilon$ can amount to a 10 \% reduction, 
that the large Dirac Yukawa coupling can produce a 10 or 20~\% 
effect \cite{Vissani:1994fy}, similar to what one can estimate 
for the change due to the full three flavor analysis.

3) Last but not least there is an important relation between 
the quark mixing $V_{cb}$ and the atmospheric mixing angle. 
Eq.~(\ref{quaglia}) shows that the main part of the 
(unphysical) up- and down-quark rotations are the same; thus, the 
quark mixing is found to be:

\begin{equation}
|V_{cb}|= |\ {\rm Re}\xi -
i\cos2\theta_A\ {\rm Im}\xi  |+ {\cal O}(|\epsilon^2|)
\end{equation} 

\noindent
where $\xi=\cos2\theta_A\  (\epsilon_D-\epsilon_U)$.
This equation demonstrates the successful coexistence of small and 
large mixing angles. In order for it to work quantitatively, 
$|\cos{2\theta_A}|$ should be as large as possible, i.e.~$\theta_A$ 
should be as far as possible from the maximal value $45^\circ$. 
Strictly speaking, even this would not be sufficient if this 
prediction is taken at its face value. However the neglected effects 
from the first generation and the loops prevent us from sentencing 
this prediction and this model. 

The analysis we presented above could be in principle changed by the 
two loop corrections in the Yukawa sector \cite{Bajc:2004hr}. In 
the model I this consists only in the renormalization of the original 
couplings, while in model II it could generate an effective coupling 
of the fermions to a one-index object (${\bf 10_H^{eff}}$) such as for 
example 

\begin{equation}
{\bf 16_F}\frac{{\bf 210_H}{\bf\overline{126}_H}}{M}{\bf 16_F}\;,\;
{\bf 16_F}\frac{{\bf 45_H}{\bf 120_H}}{M}{\bf 16_F}\;,\;...
\end{equation}

\noindent
Such terms, even if present, are negligible in the present study of the 
two generation case, but they should be taken into account in the full 
analysis of the three generations.

\section{Symmetry breaking patterns and neutrino mass}

As argued already in the introduction, SO(10) GUT works perfectly 
well without invoking supersymmetry. It is true that supersymmetry 
leads naturally to the unification of gauge couplings, but the same 
effect can be equally achieved with left-right symmetry as an
intermediate scale. This is precisely what happens in SO(10). In the 
over-constrained models discussed in this paper, the Dirac neutrino 
Yukawa couplings are not arbitrary. Thus one must make sure that the 
pattern of intermediate mass scale is consistent with a see-saw mechanism
for neutrino masses. More precisely, the $B-L$-breaking scale responsible 
for right-handed neutrino masses cannot be too low. On the other hand, 
this scale, strictly speaking, cannot be predicted by the renormalization 
group study of the unification constraints. The problem is that the 
right-handed neutrinos and the Higgs scalar responsible for $B-L$ 
breaking are Standard Model singlets, and thus have almost no impact 
(zero impact at one-loop) on the running. Fortunately, we know that 
the $B-L$ breaking scale must be below SU(5) breaking, since the 
couplings do not unify in the Standard Model. Better to say, 
$M_{B-L} \leq M_R$, the scale of SU(2)$_R$ breaking, and hence one 
must make sure that $M_R$ is large enough. This, together with proton 
decay constraints, will allow us to select between a large number of 
possible patterns of symmetry breaking.

Our task is simplified by the exhaustive study of symmetry breaking 
in the literature, in particular the careful two-loop level calculations 
of Ref.~\cite{Deshpande:1992em}. Recall, though, that the $(2,2,15)$ 
field must lie below the GUT scale as discussed in sect.~\ref{whois} 
and although its impact on the running is very tiny, it must be included.   

The lower limit on $M_R$ stems from the heaviest neutrino mass

\begin{equation}
m_\nu \geq \frac{m_t^2}{M_R}\;,
\end{equation}

\noindent
which gives $M_R \geq 10^{13}$ GeV or so. One can now turn 
to the useful table of Ref.~\cite{Deshpande:1992em}, where the most 
general patterns of SO(10) symmetry breaking with two intermediate 
scales consistent with proton decay limits are presented. (Notice 
that the models with subscript 'b' in the table utilize ${\bf 16_H}$ 
in place of ${\bf \overline{126}_H}$ to break the SU(2)$_R$
symmetry, and thus are not relevant for our discussion.)

The above limit on $M_R$ immediately rules out a number of the 
remaining possibilities; the most promising candidates are those 
with an intermediate SU(2)$_L\times $SU(2)$_R\times$ SU(4)$_C\times $P  
symmetry breaking scale (that is, PS group with unbroken 
parity). This is the case in which the breaking at the large scale is 
achieved by a Pati-Salam parity even singlet, for example contained in 
${\bf 54_H}$. In the searching for a realistic symmetry breaking 
pattern one does not need to stick to the global minimum of the potential
as in \cite{Acampora:1994rh}, but a local meta-stable minimum with 
a long enough lifetime will do the job as well. 
It has to be stressed however, that a big uncertainty is implicit 
in all models with complicated or unspecified Higgs sector, due to 
possibly large and uncontrolled threshold corrections \cite{Dixit:1989ff}. 

In any case, the nature of the GUT Higgs and the pattern of symmetry 
breaking will also enter into the fitting of fermion masses, since 
they determine the decomposition of the light (fine-tuned) Higgs 
doublet (e.g., they provide relations among the parameters 
$c_{1}$, $c_{2}$, $c_{3}$, $c_{4}$ in eq.~(\ref{lec})).
This point is often overlooked but it is essential in the 
final test of the theory. At this point, for us it is reassuring 
that both the pattern of symmetry breaking and the nature of Yukawa 
interactions allow for a possibly realistic, predictive minimal model 
of non-supersymmetric SO(10). 

\section{Summary and outlook}

The recent years have witnessed an in-depth study of supersymmetric 
SO(10) grand unification based on the renormalizable see-saw 
mechanism. What has emerged is a possibly realistic picture for the 
unification of matter and forces with a predictive pattern for neutrino 
masses and mixings. The crucial point is that the SO(10) symmetry may 
be sufficient by itself, without the need for any additional physics. 
While the theory has a number of appealing features, it may be killed 
by its main ingredient: there may not be low-energy supersymmetry. 
It may be partially or completely broken. A nice example of partial 
breaking is the so-called split supersymmetry with light higgsinos and 
gauginos and heavy scalars. This picture allows for the interesting 
possibility of a radiative see-saw mechanism for neutrino masses, 
and another simple predictive version of the SO(10) theory.  

Since we know nothing about the existence of supersymmetry or the 
nature of its breaking, it is mandatory to study the non-supersymmetric 
version, as a part of the search for {\it the} SO(10) GUT. This was 
the scope of our paper. We have identified two potentially realistic, 
predictive Yukawa structures for the case of the renormalizable see-saw 
mechanism, based on a ${\bf \overline{126}_H}$. This choice is motivated 
by the fact that the alternative radiative see-saw seems to favor 
split supersymmetry \cite{Bajc:2005aq}. We have focused on the 
renormalizable version simply in order to be predictive, without 
invoking unknown physics. 

The models require adding ${\bf 10_H}$ or ${\bf 120_H}$ fields. The 
latter is particularly interesting, due to the small number of Yukawa 
couplings. Both models seem to require adding 
U(1)$_{PQ}$. While this may be appealing since it provides the axion 
as a  dark matter candidate, it is against 
the spirit of sticking to pure grand unification.

A number of issues must be addressed in order 
to construct a fully realistic theory. The first task, 
as we repeatedly argued, is a complete 
three generation numerical study. This also includes 
the construction of the minimal GUT Higgs sector and the study of its
impact on the fermion masses and mixings. For a successful model, if 
any, one must study in turn the proton decay predictions, and in 
particular, the branching ratios that are calculable in the 
over-constrained theories discussed here. This is a less urgent (but 
equally important) task simply due to the lack of  experimental data.  
Beside proton decay, the other generic feature of grand unification is the 
existence of magnetic monopoles which brings along the so-called monopole 
problem due to the over-production in the early universe. While there is 
always the possibility of the inflation solution, it is worth recalling 
that in non supersymmetric theories there are other interesting ways out 
of this impasse. These are for example the symmetry non-restoration 
at high temperature \cite{Dvali:1995cj} and the possibility that unstable domain 
walls sweep the monopole away \cite{Dvali:1997sa}. In principle, 
either of these alternative solutions can provide further 
constraints on the parameters of the theory. Yet another constraint
comes from leptogenesis, which finds its natural role in SO(10) with seesaw 
mechanism.
 
\acknowledgments
This work was completed during the 2005 Gran Sasso Summer 
Institute, and we wish to thank Z.~Berezhiani for  
organizing an excellent meeting  and for his warm hospitality. 
The work of G.S.\ was supported in part by European Commission 
under the RTN contract MRTN-CT-2004-503369; 
the work of B.B.\ by the Ministry of Education, 
Science and Sport of the Republic of Slovenia: the work of 
A.M.\ by CDCHT-ULA project No.\ C-1244-04-05-B.  B.B.\ thanks 
Michal Malinsk\'y and Miha Nemev\v{s}ek for discussion.


\begin{thebibliography}{999}

\bibitem{Fritzsch:nn}
H. Georgi, {\em In Coral Gables 1979 Proceeding, Theory and experiments
in high energy physics,} New York 1975, 329 and
H.~Fritzsch and P.~Minkowski,
Annals Phys.\  {\bf 93} (1975) 193.

\bibitem{Pati:1974yy}
  J.~C.~Pati and A.~Salam,
  Phys.\ Rev.\ D {\bf 10} (1974) 275.

\bibitem{leftright}  R.~N.~Mohapatra and J.~C.~Pati,
  Phys.\ Rev.\ D {\bf 11} (1975) 2558;
  G.~Senjanovi\' c and R.~N.~Mohapatra,
  Phys.\ Rev.\ D {\bf 12} (1975) 1502;
G.~Senjanovi\' c,
  Nucl.\ Phys.\ B {\bf 153} (1979) 334.

\bibitem{Minkowski:1977sc}
P.~Minkowski,
Phys.\ Lett.\ B {\bf 67} (1977) 421; 
T.~Yanagida, proceedings of the {\em Workshop on Unified Theories 
and Baryon Number in the Universe}, Tsukuba, 1979, eds. 
A. Sawada, A. Sugamoto; 
S.~Glashow, in {\em Cargese 1979, Proceedings, Quarks and Leptons}
(1979); 
M.~Gell-Mann, P.~Ramond, R.~Slansky, proceedings of the
{\em Supergravity Stony Brook Workshop}, New York, 1979, 
eds. P. Van Niewenhuizen, D. Freeman; 
R.~Mohapatra, G.~Senjanovi\' c,
Phys.Rev.Lett. {\bf 44} (1980) 912.

\bibitem{Dvali:2003br}  G.~Dvali and A.~Vilenkin,
  Phys.\ Rev.\ D {\bf 70} (2004) 063501
  [arXiv:hep-th/0304043];
 G.~Dvali,
  [arXiv:hep-th/0410286].
 
\bibitem{Langacker:1991an}
  P.~Langacker and M.~x.~Luo,
  Phys.\ Rev.\ D {\bf 44} (1991) 817.

\bibitem{susyunif}
S.~Dimopoulos, S.~Raby, F.~Wilczek,
Phys.\ Rev.\ D {\bf 24} (1981) 1681.
L.E.~Ib\'a\~nez, G.G.~Ross,
Phys.\ Lett.\ B {\bf 105} (1981) 439.
M.B.~Einhorn, D.R.~Jones,
Nucl.\ Phys.\ B {\bf 196} (1982) 475.
W.~Marciano, G.~Senjanovi\' c,
Phys.Rev.D {\bf 25} (1982) 3092.

\bibitem{Aulakh:2003kg}
  C.~S.~Aulakh, B.~Bajc, A.~Melfo, G.~Senjanovi\' c and F.~Vissani,
  Phys.\ Lett.\ B {\bf 588}, 196 (2004)
  [arXiv:hep-ph/0306242].

\bibitem{Senjanovic:2005sf}
  G.~Senjanovi\' c,
  arXiv:hep-ph/0501244.

\bibitem{Deshpande:1992em}
  N.~G.~Deshpande, E.~Keith and P.~B.~Pal,
  Phys.\ Rev.\ D {\bf 47}, 2892 (1993)
  [arXiv:hep-ph/9211232].

\bibitem{Acampora:1994rh}
F.~Acampora, G.~Amelino-Camelia, F.~Buccella, O.~Pi\-san\-ti, 
L.~Rosa and T.~Tuzi, Nuovo Cim.\ A {\bf 108} (1995) 375
[arXiv:hep-ph/9405332].

\bibitem{Peccei:1977hh}
  R.~D.~Peccei and H.~R.~Quinn,
  Phys.\ Rev.\ Lett.\  {\bf 38} (1977) 1440.

\bibitem{Bajc:2004xe}
B.~Bajc, A.~Melfo, G.~Senjanovi\' c and F.~Vissani,
  Phys.\ Rev.\ D {\bf 70} (2004) 035007
  [arXiv:hep-ph/0402122].

\bibitem{Aulakh:2002zr}
C.~S.~Aulakh and A.~Girdhar,
  Int.\ J.\ Mod.\ Phys.\ A {\bf 20} (2005) 865
  [arXiv:hep-ph/0204097].

\bibitem{Fukuyama:2004ps}
T.~Fukuyama, A.~Ilakovac, T.~Kikuchi, S.~Meljanac and N.~Okada,
  J.\ Math.\ Phys.\  {\bf 46}, 033505 (2005)  [arXiv:hep-ph/0405300].

\bibitem{Matsuda:2000zp}
  K.~Matsuda, Y.~Koide and T.~Fukuyama,
  Phys.\ Rev.\ D {\bf 64} (2001) 053015
  [arXiv:hep-ph/0010026].

\bibitem{Matsuda:2001bg}
  K.~Matsuda, Y.~Koide, T.~Fukuyama and H.~Nishiura,
  Phys.\ Rev.\ D {\bf 65} (2002) 033008
  [Erratum-ibid.\ D {\bf 65} (2002) 079904]
  [arXiv:hep-ph/0108202]; 
  T.~Fukuyama and N.~Okada,
  JHEP {\bf 0211} (2002) 011
  [arXiv:hep-ph/0205066].

\bibitem{Goh:2003sy}
  H.~S.~Goh, R.~N.~Mohapatra and S.~P.~Ng,
  Phys.\ Lett.\ B {\bf 570} (2003) 215
  [arXiv:hep-ph/0303055]; 
  H.~S.~Goh, R.~N.~Mohapatra and S.~P.~Ng,
  Phys.\ Rev.\ D {\bf 68} (2003) 115008
  [arXiv:hep-ph/0308197]; 
  B.~Dutta, Y.~Mimura and R.~N.~Mohapatra,
  Phys.\ Rev.\ D {\bf 69} (2004) 115014
  [arXiv:hep-ph/0402113].

\bibitem{Babu:2005ia}
  K.~S.~Babu and C.~Macesanu,
  arXiv:hep-ph/0505200.

\bibitem{Bertolini:2005qb}
  S.~Bertolini and M.~Malinsky,
  arXiv:hep-ph/0504241.

\bibitem{Bajc:2002iw}
B.~Bajc, G.~Senjanovi\' c, F.~Vissani,
Phys.\ Rev.\ Lett.\  {\bf 90} (2003) 051802 [hep-ph/0210207], 
and hep-ph/0110310.

\bibitem{Bajc:2004fj}
  B.~Bajc, G.~Senjanovi\' c and F.~Vissani,
  Phys.\ Rev.\ D {\bf 70} (2004) 093002
  [arXiv:hep-ph/0402140].

\bibitem{rparity}
R.~N.~Mohapatra,
{\em Phys.\ Rev.\ } {\bf D 34}, 3457 (1986).
A.~Font, L.~E.~Ib\'a\~nez and F.~Quevedo,
{\em Phys.\ Lett.\ } {\bf B228}, 79 (1989).   
S.~P.~Martin,
{\em Phys.\ Rev.\ } {\bf D46}, 2769 (1992).

\bibitem{Aulakh:1997ba}
C.S.~Aulakh, K.~Benakli, G.~Senjanovi\'c,
Phys.\ Rev.\ Lett.\  {\bf 79} (1997) 2188.
C.~S.~Aulakh, A.~Melfo and G.~Senjanovi\'c, 
Phys.\ Rev.\ D {\bf 57}, 4174 (1998);
[arXiv:hep-ph/9707256].
C.~S.~Aulakh, A.~Melfo, A.~Ra\v{s}in and G.~Senjanovi\'c,
  Phys.\ Rev.\ D {\bf 58}, 115007 (1998)
  [arXiv:hep-ph/9712551].
C.~S.~Aulakh, A.~Melfo, A.~Ra\v{s}in and G.~Senjanovi\'c,
Phys.\ Lett.\ B {\bf 459} (1999) 557.
[arXiv:hep-ph/9902409].


\bibitem{Bajc:2005aq}
  B.~Bajc and G.~Senjanovi\' c,
  arXiv:hep-ph/0507169.
 
\bibitem{Lazarides:1980nt}
  G.~Lazarides, Q.~Shafi and C.~Wetterich,
  Nucl.\ Phys.\ B {\bf 181} (1981) 287.

\bibitem{Mohapatra:1980yp}
  R.~N.~Mohapatra and G.~Senjanovi\' c,
  Phys.\ Rev.\ D {\bf 23} (1981) 165.
 
\bibitem{Babu:1992ia}
  K.~S.~Babu and R.~N.~Mohapatra,
  Phys.\ Rev.\ Lett.\  {\bf 70} (1993) 2845
  [arXiv:hep-ph/9209215].
 L. Lavoura, Phys.\ Rev.\ D{\bf 48} (1993) 5440  [arXiv:hep-ph/9306297].
 
\bibitem{Arason:1992eb}
H.~Arason, D.~J.~Castano, E.~J.~Piard and P.~Ramond,
  Phys.\ Rev.\ D {\bf 47} (1993) 232
  [arXiv:hep-ph/9204225].

\bibitem{axions}  S.~Weinberg,
  Phys.\ Rev.\ Lett.\  {\bf 40} (1978) 223;
F.~Wilczek,
  Phys.\ Rev.\ Lett.\  {\bf 40} (1978) 279.

\bibitem{Mohapatra:1982tc}
  R.~N.~Mohapatra and G.~Senjanovi\' c,
  Z.\ Phys.\ C {\bf 17} (1983) 53.

\bibitem{Kim:1986ax}  
  J.~E.~Kim,
  Phys.\ Rept.\  {\bf 150} (1987) 1.


\bibitem{Holman:1982tb}
  R.~Holman, G.~Lazarides and Q.~Shafi,
  Phys.\ Rev.\ D {\bf 27} (1983) 995.

\bibitem{Wise:1981ry}
  M.~B.~Wise, H.~Georgi and S.~L.~Glashow,
  Phys.\ Rev.\ Lett.\  {\bf 47} (1981) 402.

\bibitem{Ellis:1978hq}
J.~R.~Ellis and M.~K.~Gaillard,
Nucl.\ Phys.\ B {\bf 150} (1979) 141.

\bibitem{Beg:1978mt}
  M.~A.~B.~Beg and H.~S.~B.~Tsao,
  Phys.\ Rev.\ Lett.\  {\bf 41} (1978) 278.

\bibitem{Mohapatra:1978fy}
  R.~N.~Mohapatra and G.~Senjanovi\' c,
  Phys.\ Lett.\ B {\bf 79}, 283 (1978).

\bibitem{Georgi:1978xz}
  H.~Georgi,
  Hadronic J.\  {\bf 1} (1978) 155.
  
\bibitem{Witten:1979nr}
E.~Witten,
Phys.\ Lett.\ B {\bf 91} (1980) 81.

\bibitem{Arkani-Hamed:2004fb}
N.~Arkani-Hamed and S.~Dimopoulos, arXiv:hep-th/0405159;
  G.~F.~Giudice and A.~Romanino,
Nucl.\ Phys.\ B {\bf 699} (2004) 65 [arXiv:hep-ph/0406088].
N.~Arkani-Hamed, S.~Dimopoulos, G.~F.~Giudice and A.~Romanino,
arXiv:hep-ph/0409232.

\bibitem{Bajc:2004hr}
  B.~Bajc and G.~Senjanovi\' c,
  Phys.\ Lett.\ B {\bf 610} (2005) 80
  [arXiv:hep-ph/0411193].

\bibitem{Nath:2001uw}
  P.~Nath and R.~M.~Syed,
  Phys.\ Lett.\ B {\bf 506} (2001) 68
  [Erratum-ibid.\ B {\bf 508} (2001) 216]
  [arXiv:hep-ph/0103165].

\bibitem{Strumia:2005tc} 
  For a recent analysis and references, see  A.~Strumia and F.~Vissani,
  Nucl.\ Phys.\ B {\bf 726} (2005) 294
  [arXiv:hep-ph/0503246].
  
\bibitem{Vissani:1994fy}
  F.~Vissani and A.~Y.~Smirnov,
  Phys.\ Lett.\ B {\bf 341} (1994) 173
  [arXiv:hep-ph/9405399].

\bibitem{Dixit:1989ff}
  V.~V.~Dixit and M.~Sher,
  Phys.\ Rev.\ D {\bf 40} (1989) 3765; 
C.~S.~Aulakh and A.~Girdhar,
Nucl.\ Phys.\ B {\bf 711} (2005) 275
[arXiv:hep-ph/0405074].

\bibitem{Dvali:1995cj}
  G.~R.~Dvali, A.~Melfo and G.~Senjanovi\' c,
  Phys.\ Rev.\ Lett.\  {\bf 75},  4559 (1995)
  [arXiv:hep-ph/9507230].

\bibitem{Dvali:1997sa}
G.~R.~Dvali, H.~Liu and T.~Vachaspati,
  Phys.\ Rev.\ Lett.\  {\bf 80}, 2281 (1998)
  [arXiv:hep-ph/9710301].

\end{thebibliography}
\end{document}